\documentclass[aps,prl,reprint,superscriptaddress]{revtex4-2}

\usepackage{amsmath}
\usepackage{graphicx} 
\usepackage{braket}
\usepackage{threeparttable}
\usepackage[T1,OT1]{fontenc} 
\usepackage{float}
\usepackage{amssymb,bm}
\usepackage{orcidlink}
\usepackage{xcolor}
\usepackage[normalem]{ulem}

\makeatletter
\DeclareMathVersion{SMnormal}
\DeclareMathVersion{SMbold}
\SetSymbolFont{operators}   {SMnormal}{OT1}{lmr} {m}{n}
\SetSymbolFont{letters}     {SMnormal}{OML}{lmm} {m}{it}
\SetSymbolFont{symbols}     {SMnormal}{OMS}{lmsy}{m}{n}
\SetSymbolFont{largesymbols}{SMnormal}{OMX}{lmex}{m}{n}
\SetSymbolFont{operators}   {SMbold}  {OT1}{lmr} {bx}{n}
\SetSymbolFont{letters}     {SMbold}  {OML}{lmm} {b}{it}
\SetSymbolFont{symbols}     {SMbold}  {OMS}{lmsy}{b}{n}
\SetSymbolFont{largesymbols}{SMbold}  {OMX}{lmex}{m}{n}
\SetMathAlphabet{\mathbf}{SMnormal}{OT1}{lmr}{bx}{n}
\SetMathAlphabet{\mathsf}{SMnormal}{OT1}{lmss}{m}{n}
\SetMathAlphabet{\mathit}{SMnormal}{OT1}{lmr}{m}{it}
\SetMathAlphabet{\mathtt}{SMnormal}{OT1}{lmtt}{m}{n}
\SetMathAlphabet{\mathbf}{SMbold}  {OT1}{lmr}{bx}{n}
\SetMathAlphabet{\mathsf}{SMbold}  {OT1}{lmss}{bx}{n}
\SetMathAlphabet{\mathit}{SMbold}  {OT1}{lmr}{bx}{it}
\SetMathAlphabet{\mathtt}{SMbold}  {OT1}{lmtt}{m}{n}
\let\SM@author\author
\let\SM@affiliation\affiliation
\let\SM@email\email
\let\SM@thanks\thanks
\let\SM@and\and
\let\SM@abstract\abstract
\let\SM@endabstract\endabstract
\makeatother

\begin{document}

\title{Spin-Selective Hadron Spectroscopy via Azimuthal Anisotropies from Entanglement-Enabled Spin Interference}

\author{Samuel Corey\,\orcidlink{0009-0002-3857-2015}}
\email{corey.90@osu.edu}
\affiliation{Department of Physics, The Ohio State University, Columbus, Ohio 43210, USA}

\author{James Daniel Brandenburg\,\orcidlink{0000-0002-6327-5947}}
\email{brandenburg.89@osu.edu}
\affiliation{Department of Physics, The Ohio State University, Columbus, Ohio 43210, USA}

\author{Shuo Lin}
\email{shuolin@sdu.edu.cn}
\affiliation{Key Laboratory of Particle Physics and Particle Irradiation (MOE), Institute of Frontier and Interdisciplinary Science, Shandong University, Qingdao, Shandong 266237, China}

\author{Jian Zhou}
\email{jzhou@sdu.edu.cn}
\affiliation{Key Laboratory of Particle Physics and Particle Irradiation (MOE), Institute of Frontier and Interdisciplinary Science, Shandong University, Qingdao, Shandong 266237, China}
\affiliation{Southern Center for Nuclear-Science Theory (SCNT), Institute of Modern Physics, Chinese Academy of Sciences, Huizhou, Guangdong 516000, China}

\author{Ya-jin Zhou}
\email{zhouyj@sdu.edu.cn}
\affiliation{Key Laboratory of Particle Physics and Particle Irradiation (MOE), Institute of Frontier and Interdisciplinary Science, Shandong University, Qingdao, Shandong 266237, China}

\begin{abstract}
    The $\pi^+\pi^-$ invariant mass spectrum above the $\rho^0(770)$ is rich with broad, overlapping resonances. Disentangling them, whether in photoproduction, ultra-peripheral heavy-ion collisions, or electroproduction, is a longstanding challenge for conventional partial-wave analysis. We show that the recently observed entanglement-enabled spin-interference effect in ultra-peripheral collisions provides a quantum mechanical filter that resolves this ambiguity: the angular harmonics $A_n$ of the $\cos(n\Delta\phi)$ asymmetry, which are governed by selection rules in the spin of the interfering states.
    Specifically, overlap between two distinct spin 1 amplitudes leads to interference that appears in $A_2$ alone. Utilizing ALICE data in the $1.0$--$1.4\,\mathrm{GeV} \; c^{-2}$ region, we demonstrate that two physically distinct hypotheses---an additional spin-1 $\rho'(1450)$ (produced via photonuclear interactions) versus a spin-2 (photon-photon) $f_2(1270)$ state---fit the invariant mass spectrum equally well but predict different $A_n$: vanishing $A_1$ and $A_3$ for spin 1, versus phase-sensitive enhancements for spin 2. Measuring these harmonics exposes interference that cancels in the azimuthally integrated mass spectrum, constraining resonant and nonresonant contributions and offering a complementary probe of low energy hadron dynamics.
\end{abstract}

\maketitle

\textit{Introduction---}Ultra-peripheral heavy-ion collisions (UPCs) at collider facilities have emerged as a pristine environment for studying high-energy photon-induced interactions. Because the colliding nuclei pass each other at impact parameters larger than the sum of their radii, strong hadronic interactions are heavily suppressed, while the nuclei's intense electromagnetic fields act as a dense flux of quasi-real, linearly polarized photons~\cite{Baltz:2007kq,Klein:2020fmr, Brandenburg:2025one,Li:2019sin,Li:2019yzy,Jia:2024hen,Wang:2022gkd,Lin:2022flv}. This environment enables high-precision studies of both exclusive vector meson photoproduction and photon-photon ($\gamma\gamma$) interactions \cite{star21, ATLAS:2018, ATLAS:2019azn, ATLAS:2023, CMS:2019, CMS:2021}. The exclusive photoproduction of $\pi^+\pi^-$ pairs is very well studied: a deep experimental record stretching from bubble chamber experiments \cite{Cambridge:1965, SLAC:1969, ABBHHM:1968}, through HERA $ep$ photoproduction~\cite{ZEUS:1995bfs,ZEUS:1997rof,H1:2006ogl}, to high-statistics UPC measurements at RHIC~\cite{STAR:2002, STAR:2011, STAR:2017enh} and the LHC~\cite{ALICE:2015nbw,CMS:2019awk,lhcb2pi} has established the dominance of the $\rho^0(770)$. Similarly, detailed study of the $\gamma\gamma \to \pi^+ \pi^-$ process with electron-positron collider experiments at SLAC~\cite{MarkII:1990} and KEKB~\cite{Belle:2007zza} have determined dominance of the $f_2(1270)$.

Below $1\;\mathrm{GeV}\;c^{-2}$, the unique structure of the UPC dipion invariant mass spectrum has been described in great detail. The Drell-Söding mechanism, where the $\rho^0$ amplitude interferes with a $\pi^+\pi^-$ continuum, describes the apparent shift of the $\rho^0$ mass, and presence of $\omega(782)\to\pi^+\pi^-$ account for another small feature~\cite{PhysRevLett.5.342, soding1966, STAR:2017enh}. Above $1\;\mathrm{GeV}\;c^{-2}$, however, the spectrum becomes intractable due to the many overlapping resonances. In this region, standard partial wave analysis techniques cannot cleanly disentangle, e.g., the spin-2 $f_2(1270)$ from the spin-1 $\rho'(1450)$ due to the overwhelming contributions of other broad spin-1 states\cite{Hammoud:2020aqi}.

The recent observation of entanglement-enabled spin interference (EESI) in UPCs at STAR  and ALICE provides a new handle on this problem ~\cite{STAR:2022wfe,STAR:2026vgm,Brandenburg:2024ksp, STAR:2025wpi, ALICE:2024ife}. In UPCs, the linear polarization of photons and two-way ambiguity between photon emitter and target, give rise to a $\cos(2\Delta\phi)$ modulation between the daughter $\pi^+$ and the pair transverse momentum~\cite{Zhou:2022twz}. The factor 2 is a direct result of the angular momentum quantum numbers involved in photonuclear production. Additional $\cos(\Delta\phi)$ and $\cos(3\Delta\phi)$ are predicted as a result of interference between photonuclear and $\gamma\gamma$ channels, which carry 1 and 2 units of angular momentum respectively \cite{hagiwara21}. Here, we extend this interference approach to resonant production, using UPCs as a natural interferometer for weaker resonances.

In this Letter, we demonstrate that EESI provides a method to isolate interference terms according to their spin structure despite these terms vanishing in the azimuthally-integrated cross section. This property may allow experimentalists to distinguish the spin-2 $f_2(1270)$ from the spin-1 $\rho'(1450)$, which has been hitherto impossible, as well as study resonant and nonresonant $\gamma\gamma$ contributions despite them being obscured by wide photonuclear contributions ~\cite{Boglione:1998rw,Garcia-Martin:2011iqs, ATLAS:2019azn}. Related azimuthal observables have been used to probe
two-photon helicity amplitudes and relative phases at $e^+e^-$
colliders~\cite{Jia:2024xzx}.
This mechanism differs from the spin transfer in multibody decay
chains studied concurrently in Ref.~\cite{Zhang:2026pnd}. The remainder of the Letter develops the formalism, presents fits and predictions, and discusses the experimental and theoretical reach of the technique.

\label{sec:method}

\textit{Methodology---}Ultra-peripheral collisions, whether in the photonuclear or photon-photon channel, are mediated by quasi-real, linearly polarized photons \cite{star21, Brandenburg:2023,Shao:2023zge}. In the Equivalent Photon Approximation (EPA), these photons are manifest from the highly Lorentz-contracted electromagnetic fields of the relativistic nuclei \cite{Budnev:1975}. The geometry of this field dictates that the photons are predominantly traveling along the ion-moving direction with only very small transverse momentum oriented along the direction of the electric field \cite{Brandenburg:2023}.  Complete or partial s-channel helicity conservation, which has been measured in both $ep$ elastic photoproduction and in UPCs, creates a correlation between the helicity of the photoproduced vector meson and that of the initiating photon \cite{ZEUS:1995bfs, Zeus:1996, STAR:2009}.In the spin-1 and spin-2 cases, the azimuthal decay direction is correlated to the polarization direction of the resonant state. These correlations are not observable in asymmetric $ep$ collisions, but are realized by the interference terms in UPCs where negative contributions to the cross section are generated \cite{xing20, Brandenburg:2024ksp}.

This correlation manifests as an anisotropy in the distribution of $\Delta\phi$, defined as the angle between the pair and $\pi^+$ daughter momentum in the transverse plane. Recent measurements by the STAR and ALICE collaborations \cite{STAR:2022wfe, STAR:2026vgm, STAR:2025wpi, ALICE:2024ife} have demonstrated a $\cos(2\Delta\phi)$ signal in line with predictions \cite{xing20, Brandenburg:2024ksp,Shao:2025oeb,Brandenburg:2022jgr}. In the formalism developed by a subset of the present authors in Ref.~\cite{xing20},
\begin{equation}
    \label{2phiHelicity}
    \begin{split}
        \frac{d\sigma}{dM_{\pi\pi}} \propto \int d\mathcal{P.S.}(\epsilon^{VM*}_\perp \cdot \epsilon^\gamma_\perp)(\epsilon^{VM}_\perp \cdot \epsilon'^\gamma_\perp) \\ \approx\int d\mathcal{P.S.}(\hat P_\perp \cdot \hat k^\gamma_\perp)(\hat P_\perp \cdot \hat k'^\gamma_\perp),
    \end{split}
\end{equation}

Where $\epsilon^{VM}_\perp, P_\perp, \epsilon^\gamma_\perp$, and $k_\perp$ are the vector meson polarization, $\pi^+$ momentum, photon polarization, and photon momentum vectors respectively. Primed components belong to the conjugate amplitude. From this, the $\cos(2\Delta\phi)$ modulation can be revealed explicitly in terms of the pair transverse momentum $q_\perp$:
\begin{equation}
    \begin{split}
    (\hat P_\perp \cdot \hat k_\perp)(\hat P_\perp \cdot \hat k'_\perp) =\frac12\{\hat k_\perp \cdot \hat k_\perp' + \cos(2\Delta\phi)\times \\ [2(\hat q_\perp \cdot \hat k_\perp)(\hat q_\perp \cdot \hat k'_\perp ) - \hat k_\perp \cdot \hat k_\perp']\}.
    \end{split}
\end{equation}

While the $\cos(2\Delta\phi)$ term integrates to zero due to symmetry of the phase space about $\Delta\phi=0$, the observable
\begin{equation}
    \label{AnDef}
    A_n = \frac{\int d\mathcal{P.S.} \frac{d\sigma}{d \mathcal{P.S.}} \cos(n\Delta\phi)}{\int d\mathcal{P.S.} \frac{d\sigma}{d \mathcal{P.S.}}}
\end{equation}
with $n=2$ survives.

The interference term between $\gamma A \rightarrow \rho^0 \rightarrow \pi^+ \pi^-$ and $\gamma\gamma \rightarrow \pi^+\pi^-$ continuum can be similarly expressed \cite{hagiwara21}:
\begin{align}
\frac{d\sigma_I}{dM_{\pi\pi}} &\propto \int d\mathcal{P.S.} (\hat{P}_\perp \cdot \hat{k}'_\perp) \times \nonumber\\ 
&\left[\hat{k}_\perp \cdot \hat{\Delta}_\perp - \frac{2 P_\perp^2}{P_\perp^2 + m_\pi^2} (\hat{k}_\perp \cdot \hat{P}_\perp) (\hat{\Delta}_\perp \cdot \hat{P}_\perp)\right],\label{1phiHelicity}
\end{align}

where $\Delta_\perp$ is the transverse momentum of the second photon in the $\gamma\gamma$ process. The equivalent interference between photonuclear $\rho^0$ production and the production of a tensor resonance $R$ in photon-photon collisions,
\begin{align}
\frac{d\sigma^R_{I}}{dM_{\pi\pi}} &\propto \int d\mathcal{P.S.} \, (\hat{P}_\perp \cdot \hat{k}'_\perp) \Big[ (\hat{k}_\perp \cdot \hat{\Delta}_\perp)\left(\mathcal{M}_{++}^{R} + \mathcal{M}_{+-}^{R}\right) \nonumber \\
&\quad - 2 (\hat{k}_\perp \cdot \hat{P}_\perp) (\hat{\Delta}_\perp \cdot \hat{P}_\perp) \mathcal{M}_{+-}^{R} \Big], \label{1phiHelicity_f2}
\end{align}
is very similar in structure, but depends additionally on the two different helicity contributions of a tensor resonance $R$ $\mathcal{M}_{++}^{R}$ and $\mathcal{M}_{+-}^{R}$. See Secs. S3 and S4 of the Supplemental Materials for the full derivation \cite{SM}.

Unlike eq. \eqref{2phiHelicity}, eq.s \eqref{1phiHelicity} and \eqref{1phiHelicity_f2} can be shown to have only terms proportional to $\cos(1\Delta\phi)$ and $\cos(3\Delta\phi)$, and as such $d\sigma_I/dM_{\pi\pi}$ vanishes in the integral over $\Delta\phi$; orthogonal eigenstates do not interfere at the cross section level. Instead, this interference term manifests as $\cos(\Delta\phi)$ and $\cos(3\Delta\phi)$ azimuthal anisotropies. As this result emerges from the product of three polarization vectors, any two-photon amplitude, whether resonant or not, can contribute odd harmonics through interference with $\rho^0$; the helicity structure and relative phases govern the size of each term. Invariant mass regions with non-zero $A_1$ and $A_3$ anisotropies should therefore be associated with continuum or resonant $\gamma\gamma\to\pi^+\pi^-$ production, while non-zero $A_2$ indicates the contribution of at least two photonuclear channels. 



The differential cross section in eq. \eqref{AnDef} can be expressed as a combination of Breit-Wigner resonances $\mathcal{BW_R}$ with different properties and continuum production \cite{STAR:2017enh}. As a schematic, consider the $\rho^0(770)$ and a general resonance $X$, noting that any additional channels can be included and follow the same argument. All of the angular structure containing the $\cos(n\Delta\phi)$ dependence is denoted as $f_R(\theta, \phi)$: 
\begin{align}
    \frac{d\sigma}{d\mathcal{P.S.}}&\sim|\mathcal{M}|^2 = \,|\mathcal{BW}_\rho|^2 f_\rho(\theta,\phi) + |\mathcal{BW}_X|^2 f_X(\theta,\phi) \nonumber \\
    &\quad + (\mathcal{BW}_\rho \mathcal{BW}_X^* + \mathcal{BW}^*_\rho \mathcal{BW}_X) f_{\rho \times X}(\theta,\phi).
    \label{rhoX_matrix}
\end{align}

If $X$ is spin-1, all terms in eq. \eqref{rhoX_matrix} contain two photon polarization vectors, and thus contribute to only $A_2$. However, if this resonance has $J=2$, only the $(\rho^0)^2$ term contributes to $A_2$, and the interference terms contribute to $A_1$ and $A_3$.

The key observation from eq. \eqref{rhoX_matrix} is that terms in the cross section factorize to a resonant structure in invariant mass and an angular structure. This factorization is exact in the eikonal approximation, applied to the photons, which is well justified since the momenta of the photons are almost entirely longitudinal, while the polarizations are entirely transverse. In eq. \eqref{AnDef}, the numerator picks out the cross terms with the appropriate $\cos(n\Delta\phi)$, and the denominator is simply the total cross section. What remains is the angular integral:
\begin{equation}
    \label{AnShort}
    \begin{split}
        A_n^{\rho\times X} = \frac{(\mathcal{BW}_\rho \mathcal{BW}_X^* + \mathcal{BW}^*_\rho \mathcal{BW}_X)}{\sigma}   a_n^{\rho\times X},
    \end{split}
\end{equation}
where
\begin{equation}
    \label{small_an}
    \begin{split}
        a_n^{\rho\times X} = \int d\theta d\phi f_{\rho\times X} (\theta \phi)  \cos{n\Delta\phi}.
    \end{split}
\end{equation}
The variable $a_n^{\rho\times X}$ is the strength of the $\cos(n\Delta\phi)$ modulation owing to the $\rho\times X$ term. In eq.\eqref{AnShort}, $a_n$ is independent of kinematics, similar to the decay parameter in polarized hyperon decays \cite{STAR:2017nat}. In a model with more than two production amplitudes, the additional interference terms contribute to the $A_n$ observables through this mechanism. Odd harmonics vanish when only photonuclear spin-1 amplitudes are present. Interference with a two-photon $f_2$ or continuum amplitude can generate $A_1$ and $A_3$,  with phase-dependent signs and magnitudes. To demonstrate this, we fit existing data with two models.

Two models labeled A. and B. are fit to the data. Both include the resonances $\rho^0(770)$ and $\omega(782)$, with an additional constant representing the photonuclear continuum. In model A, an additional spin-1 resonance is included to describe the feature near $1300 \; \mathrm{MeV} \; c^{-2}$. Model B replaces the spin-1 resonance with a spin-2 state and contains an additional constant representing $\gamma\gamma$ continuum.

Each resonance is described by a relativistic Breit-Wigner distribution,
\begin{equation}
    \mathcal{BW}_X =(A_X^R + iA_X^I) \frac{\sqrt{M_{\pi\pi} M_X \Gamma_{X\rightarrow\pi^+\pi^-} }}{M_{\pi\pi}^2-M_X^2 +iM_X\Gamma_X},
\end{equation}

Where $M_X$ is the mass of resonance $X$ and $A_X^R$ and $A_X^I$ are the real and imaginary amplitudes corresponding to the resonance, used in place of a phase parameter.
\begin{equation}
    \Gamma_X = \Gamma_{0,X} \frac{M_X}{M_{\pi\pi}} \left( \frac{M_{\pi\pi}^2 - 4m_\pi^2}{M_X^2 - 4m_\pi^2} \right)^{(2j+1)/2}
\end{equation}

Is the running width for a resonance of spin $j$, $\Gamma_{0,X}$ is the bare width, and $m_\pi$ is the charged pion mass. The amplitudes, $M_X$, and $\Gamma_{0,X}$ are parameters for each resonance. The fit function of model A is
\begin{equation}
    \begin{split}
        \frac{dN}{dM_{\pi\pi}} = |\mathcal{BW}_{\rho^0(770)} + \mathcal{BW}_{\omega(782)} + \\\mathcal{BW}_{\rho'(1450)} +B_{\gamma A} |^2,
    \end{split}
\end{equation}

and of model B is 
\begin{equation}
    \begin{split}
        \frac{dN}{dM_{\pi\pi}}= |\mathcal{BW}&_{\rho^0(770)} + \mathcal{BW}_{\omega(782)}  +B_{\gamma A}|^2 \\&+|\mathcal{BW}_{f_2(1270)} + B_{\gamma\gamma} |^2,
    \end{split}
\end{equation}

where $B_{\gamma A}$ and $B_{\gamma\gamma}$ are the photonuclear and $\gamma\gamma$ continua respectively. The interference between $C$-odd and $C$-even amplitudes vanishes in the invariant-mass spectrum but can contribute to odd azimuthal asymmetries.

\begin{figure*}[t]
  \centering

  \includegraphics[width=0.95\textwidth]{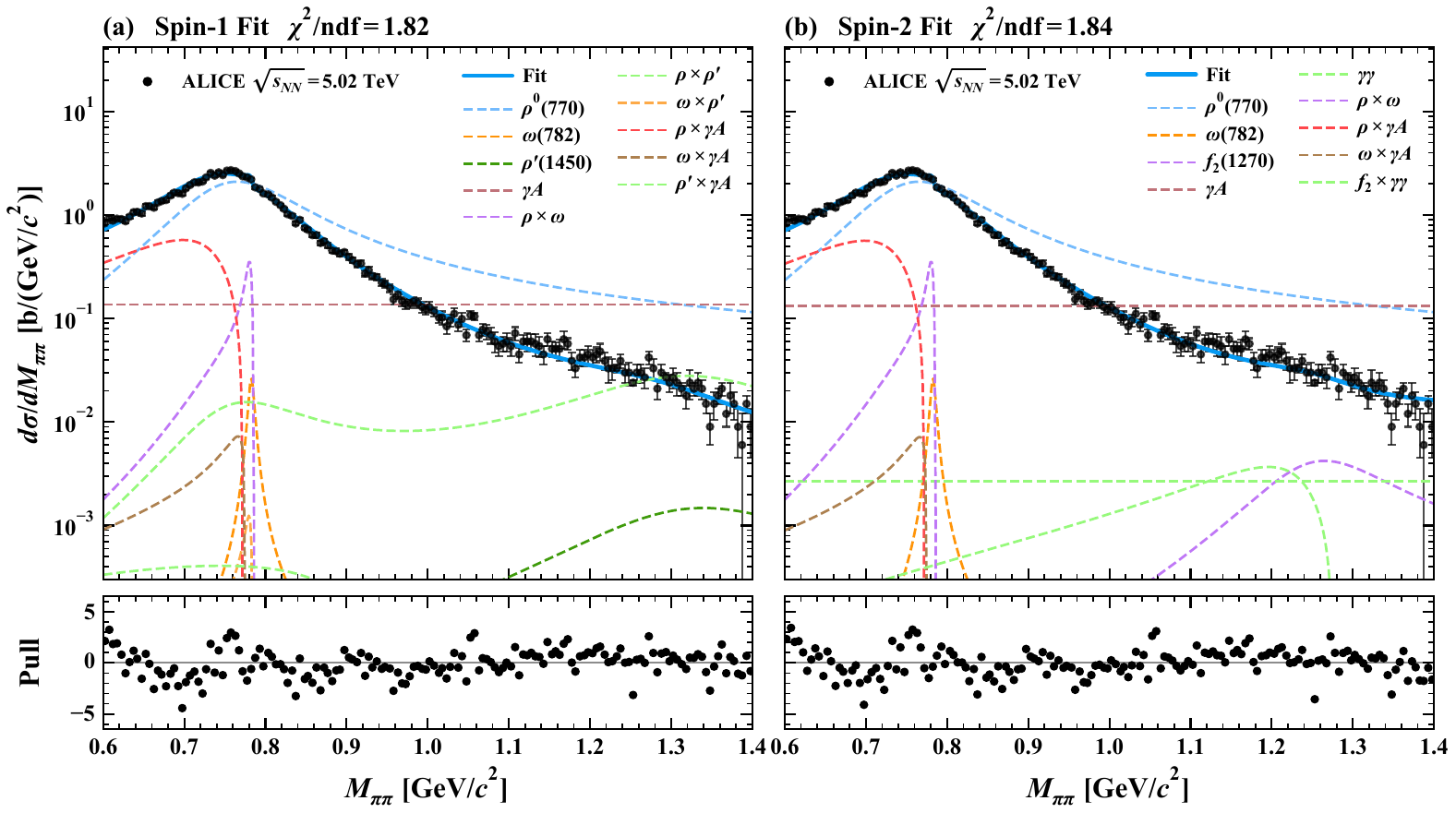}

  \caption{The $\pi^+\pi^-$ invariant mass spectrum data collected by ALICE \cite{alice2pi} fit with contributions from $\rho^0(770), \, \omega(782)$ and non-resonant continuum. In a. contributions from an additional spin-1 resonance, labeled $\rho'(1450)$ with mass and width fixed to values measured by LHCb \cite{lhcb2pi}, are included, while in b., a spin-2 resonance labeled $f_2(1270)$ with mass and width fixed to values from the Particle Data Group \cite{Navasetal2024} and a $\gamma\gamma$ continuum are considered. The phase of the $\gamma\gamma$ contributions is fixed to $\pi/2$ as the cross section is not sensitive to the interference between spin-1 and spin-2 components.} 
  \label{fig:fitComparison}
\end{figure*}

Figure \ref{fig:fitComparison} shows the results of these fits. Both are capable of describing the data, and have consistent results for the well-constrained $\rho^0, \; \omega$, and photonuclear continuum. The fits lack contribution from the wide  $\rho''(1700)$ resonance observed by both the ALICE and LHCb collaborations \cite{alice2pi, lhcb2pi}, whose Breit-Wigner shape and interference with the photonuclear continuum would contribute significantly near $1400 \; \mathrm{MeV} \; c^{-2}$. Each fit describes the data equally well and therefore cannot differentiate between the two descriptions.



The results of the fit provide the necessary information required in eq. \eqref{AnShort} to calculate the $A_n$ except for the values of $a_n$. The value of $a_2^{\rho \times \rho}$ can be extracted from the existing data \cite{STAR:2022wfe, STAR:2026vgm, ALICE:2024ife}. Lacking measurements of the odd harmonics, the values of $a_1^{\rho \times \gamma\gamma}$ and $a_3^{\rho \times \gamma\gamma}$ are acquired from theoretical calculations for the interference between $\rho^0(770)$ and the $\gamma\gamma$ continuum \cite{hagiwara21}, and $a_1^{\rho \times f_2}$ and $a_3^{\rho \times f_2}$ from Sec. S4 of the Supplemental Material \cite{SM}. In the Supplemental Material, we present the explicit
$\rho^0$--$f_2(1270)$ interference expressions and examine
the dependence on their relative production phase. In each case, a single value of $a_n$ is obtained by averaging the measured/calculated $A_n$ over the pair transverse momentum range $0.0$--$0.1 \; \mathrm{GeV}\; c^{-1}$, weighted by $\exp(- b t)$ where $b=259 \ (\mathrm{GeV/c})^{-2}$ is the radius of the lead nucleus and $t$ is the transverse momentum squared. The exponential weight approximates the transverse momentum distribution of photonuclear production. This results in the following values for $a_n$: $a_1^{\rho\times \gamma\gamma} = 0.0124$, $a_1^{\rho\times f_2} = 0.108$, $a_2 = 0.227$, $a_3^{\rho\times \gamma\gamma}  = 0.001$, and $a_3^{\rho\times f_2}  = 0.006$. Additionally, we make the simplification that $\rho'(1450)$ contributes the same $a_2$ as $\rho^0(770)$. In full generality, differences in dipole properties of the resonant state may produce different $a_n$, but the polarization-dependent structure in eq.s \eqref{2phiHelicity} and \eqref{1phiHelicity} are unchanged. Under this assumption, predictions for the observables $A_n$ can be calculated from the fit results.

\begin{figure*}[t]
  \centering

  \includegraphics[width=0.8\textwidth]{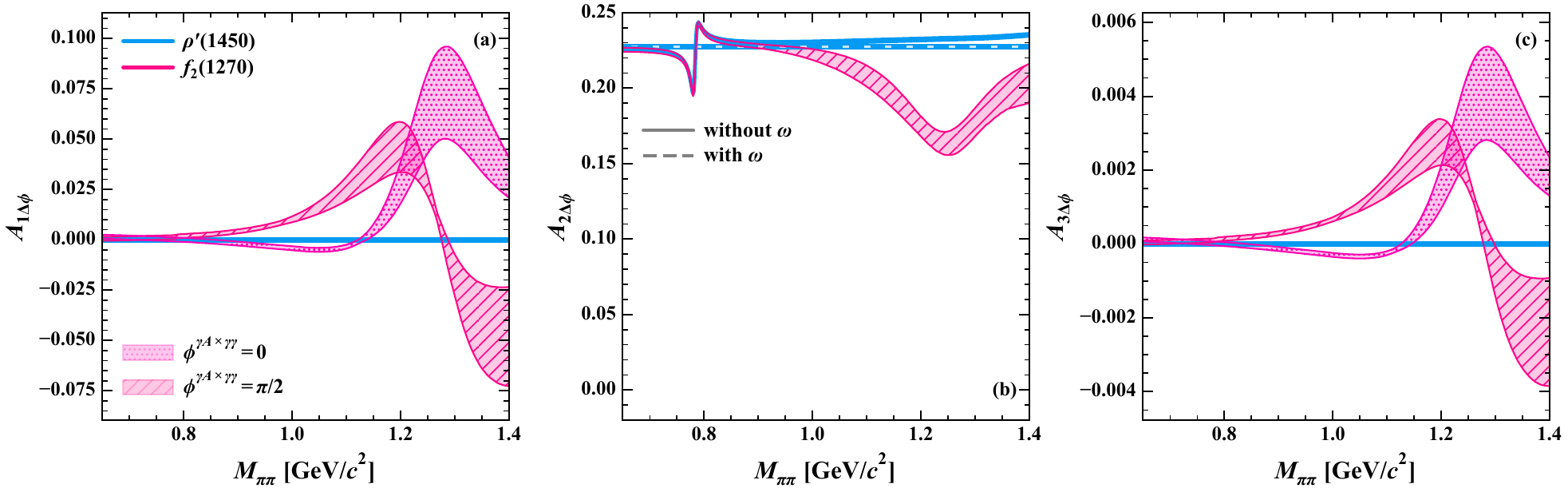}

  \caption{Predictions for the spin interference observables $A_1, \, A_2$, and $A_3$ (a., b., and c. respectively) based on the fit in fig.~\ref{fig:fitComparison}. The shaded regions represent the $1\sigma$ confidence interval. In the extraction from the $\rho'$ fit, $A_1$ and $A_3$ are 0 for all $M_{\pi\pi}$, while in the extraction from the $f_2(1270)$ fit there is a clear feature corresponding to the spin-2 resonance. The $A_1$ and $A_3$ are calculated both with the phase fixed to $0$ and to $\pi/2$ as shown in fig. \ref{fig:fitComparison}b. We provide two predictions for $A_2$ with and without contribution from the $\omega(782)$. The resultant structure suggests that measurement of $A_2$ may therefore provide a new avenue to study the nature of $\omega-\rho^0$ kinetic mixing.} 
  \label{fig:AnPrediction}
\end{figure*}

Figure \ref{fig:AnPrediction} shows these predictions for model A (blue) and model B (red). While the fit results leave the spin of the resonance near $1300 \; \mathrm{MeV} \; c^{-2}$ ambiguous, the two cases produce very different azimuthal anisotropy coefficients. $A_1$ and $A_3$ are zero in model A, but non-zero for model B at the displayed phase values, which produces a large peak at the spin-2 resonance mass. In $A_2$, model B produces a similarly large feature at the same mass, which occurs because the spin-2 resonance dilutes the $A_2$ signal produced by the spin-1 components. The ``with $\omega$'' prescription, as an approximation, assigns the common $a_2$ to the $\omega$ interference terms, whereas without it $a_2=0$. Different treatments of the $\omega$'s contribution to $A_2$ in both models produce subtle but qualitatively and quantitatively different effects that may also be identifiable in experiment. Additionally, Model A omits the $\gamma\gamma\to\pi^+\pi^-$ continuum because its inclusion greatly increases the fit uncertainties. Although adding this continuum to Model~A would generate nonzero $A_1$
and $A_3$, its odd-harmonic contribution
is substantially smaller than that from resonant
$\rho^0$--$f_2(1270)$ interference, and it lacks the characteristic $f_2(1270)$ resonant mass dependence. Thus, the
proposed discrimination remains viable.

\textit{Discussion---}The selection rules embedded in the entanglement-enabled spin-interference effect convert an intractable spectroscopy problem into a clean experimental check: when only spin-1 photonuclear amplitudes are present, odd harmonics vanish; interference with a spin-2 two-photon amplitude generates $A_1$ and $A_3$, with the sign and magnitude of each governed by the relative production phase. These harmonic patterns complement what partial-wave analysis can achieve. In the case studied here, this distinguishes the photonuclear $\rho'(1450)$ hypothesis from the $\gamma\gamma\to f_2(1270)$ hypothesis. Natural targets for the same technique include the scalar $f_0(980)/a_0(980)$ and $f_0(500)$ in $\pi\pi$ and $K\bar K$ channels, the tensor glueball candidate $f_2'(1525)$, photoproduced spin-3 or higher mesons that require contributions from the tensor pomeron, and any future measurement targeting exclusive states with mixed quantum numbers in UPCs at RHIC, the LHC, and the EIC~\cite{AbdulKhalek:2021gbh, Hagiwara4phi:2021, Bolz:2014mya, Jia:2025oak,Lin:2024mnj}. 
For $f_0(980)$, interference with $\rho^0$ makes $A_1$
sensitive to their relative phase near the $K\bar K$
threshold (Sec.~S5). The same Flatt\'e parameters also
enter BESIII's $X(2370)\to f_0(980)\eta'$ analysis; UPC measurements could therefore
help reduce scalar-amplitude uncertainty in tests of the
proposed glueball-dominant interpretation
~\cite{BESIII:2023wfi,BESIII:2026mvn}.
Complementary spectroscopic applications of EESI are emerging: concurrent work proposes using the spin-alignment transfer in multi-body decay chains to extract intermediate-state quantum numbers and branching ratios~\cite{Zhang:2026pnd}. In contrast, we use odd azimuthal harmonics to probe interference between photonuclear and two-photon amplitudes in the same dipion final state. We also note that our approach can be especially employed in light-ion collisions, like the recent O+O datasets collected at RHIC and the LHC, since the relative production cross section of the spin-1 vs. spin-2 states is expected to vary compared to collisions with heavy nuclei.

The most striking near-term implication is to constrain the $\gamma\gamma$ contribution in heavy-ion collisions~\cite{KlusekGawenda:2020}. This process is the cleanest hadronic two-photon reaction at low energies, providing direct empirical input to dispersive analyses and chiral perturbation theory~\cite{Hoferichter:2011wk,Pennington:2008xd,Dai:2014zta}. Because the photonuclear and $\gamma\gamma$ contributions to the dipion spectrum are not separable by mass alone, previous UPC measurements have treated the $\gamma\gamma$ contributions as negligible\cite{Klusek-Gawenda:2013rtu}. Odd harmonics provide additional sensitivity.

The current statistical reach of UPC datasets at both RHIC and the LHC, projected through Run~3 and beyond, is more than sufficient to perform this measurement. We estimate the statistical uncertainty using \(\Delta A_n\simeq 1/\sqrt{2N_{\rm }}\), which gives \(\Delta A_{1,3}\simeq 7.1\times10^{-4}\) for \(N_{\rm }=10^6\) and \(2.2\times10^{-2}\) for \(N_{\rm}=10^3\). For reference, the LHCb dataset \cite{LHCb:2025fzk} has \(N_{\rm }\approx10^6\) in the \(\rho^0\) peak bin and \(\approx10^3\) in the \(\rho''\) peak bin. With these uncertainties, a non-zero odd harmonic should be detectable. Two systematic limitations deserve emphasis. First, our calculation of the azimuthal asymmetries involves simplifying assumptions. The Supplemental Material derives the $\rho^0$--$f_2$ angular
dependence from the underlying amplitudes \cite{SM}. Extending this
treatment to the other interference channels would improve
this approximation. Second, the $\omega$ contribution to $A_2$ is currently uncertain at the level of $\sim 10\%$, but is itself a measurable observable that will inform future modeling of $\rho$-$\omega$ mixing in the angular sector. Both limitations should be regarded as opportunities for further work rather than obstacles: each tightens the connection between this technique and the broader programs of UPC photoproduction phenomenology and low-energy hadron physics.

By turning the recently observed entanglement-enabled spin-interference effect into a quantitative spectroscopic tool, this work elevates it from a quantum-mechanical curiosity to a practical instrument for future discovery. The same property that makes the daughter pions interfere---their entanglement combined with the two way ambiguity between photon emitter and target---becomes a selection rule on the state that produced them, and therefore a new dimension on which to project the long-standing problem of light hadron spectroscopy.

\section{Acknowledgments}
This work was supported in part by the U.S. Department of Energy, Office of Science, Office of Nuclear Physics, under contract number DE-SC0024189 as part of the Early Career Research Program, and by the National Natural Science Foundation of China under Grant
Nos.~12175118, 12321005, 12575091 (J.Z.), 12475084 (Y.Z.), as well as by the Postdoctoral Innovation Program
of Shandong Province under Project No.~SDCX-ZG-202603015 (S.L.), and the Natural Science Foundation of Shandong Province (Grant No. ZR2024MA012) (Y.Z.). We would also like to thank Zhangbu Xu, Wangmei Zha, and Raju Venugopalan for valuable discussion as well as members of the STAR and ALICE collaborations. 

\bibliographystyle{unsrt}
\bibliography{references}

\makeatletter
\close@column
\clearpage
\@booleanfalse\twocolumn@sw
\onecolumngrid
\onecolumn@grid@setup
\def\set@footnotewidth{\set@footnotewidth@one}%
\let\compose@footnotes\compose@footnotes@one
\def\title@column#1{\minipagefootnote@init#1\minipagefootnote@foot}%
\def\close@column{\newpage}%
\let\set@listindent\set@listindent@parshape
\let@environment{widetext}{widetext@galley}%
\def\marry@skip{\z@skip}%
\tolerance200 \pretolerance100 \emergencystretch8\p@ \hfuzz.1\p@ \vfuzz.1\p@
\setlength\textwidth{484.20988pt}%
\setlength\textheight{664.88487pt}%
\setlength\oddsidemargin{-7.22743pt}%
\setlength\evensidemargin{-7.22743pt}%
\setlength\topmargin{-48.22743pt}%
\setlength\columnwidth{\textwidth}%
\setlength\linewidth{\textwidth}%
\hsize\textwidth
\set@colht\set@vsize
\@floatplacement\@dblfloatplacement
\renewcommand\encodingdefault{T1}%
\renewcommand\rmdefault{lmr}%
\renewcommand\sfdefault{lmss}%
\renewcommand\ttdefault{lmtt}%
\DeclareRobustCommand\boldmath{\@nomath\boldmath\mathversion{SMbold}}%
\DeclareRobustCommand\unboldmath{\@nomath\unboldmath\mathversion{SMnormal}}%
\@ifundefined{expand@font@defaults}{}{\expand@font@defaults}%
\normalfont\mathversion{SMnormal}%
\setcounter{page}{1}%
\setcounter{section}{0}\setcounter{subsection}{0}%
\setcounter{subsubsection}{0}\setcounter{paragraph}{0}%
\setcounter{equation}{0}\setcounter{figure}{0}\setcounter{table}{0}%
\setcounter{footnote}{0}\setcounter{mpfootnote}{0}%
\setcounter{affil}{0}\setcounter{collab}{0}%
\renewcommand\theHsection{SM.\the\value{section}}%
\renewcommand\theHequation{SM.\the\value{equation}}%
\renewcommand\theHfigure{SM.\the\value{figure}}%
\renewcommand\theHtable{SM.\the\value{table}}%
\hypersetup{hidelinks}%
\let\author\SM@author
\let\affiliation\SM@affiliation
\let\email\SM@email
\let\thanks\SM@thanks
\let\and\SM@and
\let\abstract\SM@abstract
\let\endabstract\SM@endabstract
\let\@AF@join\@AF@join@error
\let\@FMN@list\@empty
\let\comma@space\@empty
\let\@address\@undefined
\let\AFF@opr\@undefined
\let\AFG@opr\@undefined
\let\CO@opr\@undefined
\let\maketitle\frontmatter@maketitle
\def\Hy@PageAnchorSlide{\def\thepage{SM\arabic{page}}}%
\def\frontmatter@footnotemark#1{%
 \leavevmode
 \ifhmode\edef\@x@sf{\the\spacefactor}\nobreak\fi
  \begingroup
   \hyper@linkstart {link}{SM.frontmatter.#1}%
    \csname c@\@mpfn\endcsname#1\relax
    \def\@thefnmark{\frontmatter@thefootnote}%
    \@makefnmark
   \hyper@linkend
  \endgroup
 \ifhmode\spacefactor\@x@sf\fi
 \relax
}%
\def\present@bibnote#1#2{%
 \item[%
  \textsuperscript{%
   \normalfont
   \Hy@raisedlink{\hyper@anchorstart{SM.frontmatter.#1}\hyper@anchorend}%
   \begingroup
    \csname c@\@mpfn\endcsname#1\relax
    \frontmatter@thefootnote
   \endgroup
  }%
 ]#2\par
}%
\long\def\SM@relabel#1\label#2#3\SM@end{\def\titleblock@produce{#1\label{SM@#2}#3}}%
\expandafter\SM@relabel\titleblock@produce\SM@end
\newwrite\SM@auxout
\if@filesw
 \immediate\openout\SM@auxout=supplemental_material_merged.aux\relax
 \immediate\write\SM@auxout{\relax}%
 \immediate\write\@auxout{\gdef\string\@extra@binfo{@SM}}%
\fi
\gdef\@extra@b@citeb{@SM}%
\let\SM@NAT@sort@cites\NAT@sort@cites
\long\def\NAT@sort@cites#1{%
 \let\SM@saved@auxout\@auxout\let\@auxout\SM@auxout
 \SM@NAT@sort@cites{#1}%
 \let\@auxout\SM@saved@auxout
}%
\def\bibliography@latex#1{%
 \if@filesw
  \immediate\write\SM@auxout{\string\bibdata{\zap@space#1 \@empty}}%
  \immediate\write\SM@auxout{\string\bibstyle{\@bibstyle}}%
 \fi
 \@input@{supplemental_material_merged.bbl}%
}%
\def\NAT@bibsetnum#1{\setlength{\topsep}{\z@}\NATx@bibsetnum{\ref{SM@LastBibItem}}}%
\let\@bibsetup\NAT@bibsetnum
\long\def\endthebibliography{%
 \auto@bib@innerbib
 \edef\@currentlabel{\arabic{NAT@ctr}}%
 \label{SM@LastBibItem}%
 \endNAT@thebibliography
 \aftergroup\auto@bib@empty
}%
\makeatother
\newcommand{\dd}{\mathrm{d}}
\newcommand{\ii}{\mathrm{i}}
\newcommand{\Ree}{\operatorname{Re}}
\newcommand{\Imm}{\operatorname{Im}}
\newcommand{\added}[1]{{\color{black}#1}}
\renewcommand{\thesection}{S\arabic{section}}
\renewcommand{\thesubsection}{\thesection\Alph{subsection}}
\renewcommand{\theequation}{S\arabic{equation}}
\renewcommand{\thefigure}{S\arabic{figure}}
\renewcommand{\thetable}{S\arabic{table}}

\setcounter{secnumdepth}{3}

\title{Supplemental Material for `Spin-Selective Hadron Spectroscopy via Azimuthal Anisotropies from Entanglement-Enabled Spin Interference'}

\author{Samuel Corey\,\orcidlink{0009-0002-3857-2015}}
\email{corey.90@osu.edu}
\affiliation{Department of Physics, The Ohio State University, Columbus, Ohio 43210, USA}

\author{James Daniel Brandenburg\,\orcidlink{0000-0002-6327-5947}}
\email{brandenburg.89@osu.edu}
\affiliation{Department of Physics, The Ohio State University, Columbus, Ohio 43210, USA}

\author{Shuo Lin}
\email{shuolin@sdu.edu.cn}
\affiliation{Key Laboratory of Particle Physics and Particle Irradiation (MOE), Institute of Frontier and Interdisciplinary Science, Shandong University, Qingdao, Shandong 266237, China}

\author{Jian Zhou}
\email{jzhou@sdu.edu.cn}
\affiliation{Key Laboratory of Particle Physics and Particle Irradiation (MOE), Institute of Frontier and Interdisciplinary Science, Shandong University, Qingdao, Shandong 266237, China}
\affiliation{Southern Center for Nuclear-Science Theory (SCNT), Institute of Modern Physics, Chinese Academy of Sciences, Huizhou, Guangdong 516000, China}

\author{Ya-jin Zhou}
\email{zhouyj@sdu.edu.cn}
\affiliation{Key Laboratory of Particle Physics and Particle Irradiation (MOE), Institute of Frontier and Interdisciplinary Science, Shandong University, Qingdao, Shandong 266237, China}

\maketitle


\added{This Supplemental Material provides the fit parameters underlying
Fig.~1 of the Letter, the explicit decomposition of the azimuthal harmonics,
and the full helicity-amplitude formalism and numerical predictions for
$\rho^0$--$f_2(1270)$ interference in Pb+Pb ultra-peripheral collisions (UPCs).
We further examine the sensitivity of the odd harmonics to the relative
production phase, tensor helicity structure, and off-shell form factors,
outline how $A_1$ probes the $f_0(980)$ line shape, and present baseline
predictions for $\rho^0$ interference with the nonresonant
$\gamma\gamma\to\pi^+\pi^-$ continuum.}

\begin{table}[H]
\centering
\begin{threeparttable}
\renewcommand{\arraystretch}{1.4}
\addtolength{\tabcolsep}{10pt} 
\begin{tabular}{l c c c}
\hline\hline
Parameter & Spin-1 Fit & Spin-2 Fit ($\phi=0$) & Spin-2 Fit ($\phi=\pi/2$) \\
\hline
$M_{\rho^0}$ [GeV/$c^2$]           & $0.7719 \pm 0.0017$ & $0.7719 \pm 0.0017$ & $0.7719 \pm 0.0017$ \\
$\Gamma_{\rho^0}$ [GeV]            & $0.1679 \pm 0.0028$ & $0.1679 \pm 0.0028$ & $0.1679 \pm 0.0028$ \\
$A_{\rho^0}$                       & $0.5915 \pm 0.0029$ & $0.5915 \pm 0.0029$ & $0.5915 \pm 0.0029$ \\
\hline
$M_{\omega}$ [GeV/$c^2$]           & $0.78266$ (fixed) & $0.78266$ (fixed) & $0.78266$ (fixed) \\
$\Gamma_{\omega}$ [GeV]            & $0.0086$ (fixed) & $0.0086$ (fixed) & $0.0086$ (fixed) \\
$A^R_{\omega}$                     & $0.0576 \pm 0.0244$ & $0.0576 \pm 0.0244$ & $0.0576 \pm 0.0244$ \\
$A^I_{\omega}$                     & $0.1068 \pm 0.0275$ & $0.1068 \pm 0.0275$ & $0.1068 \pm 0.0275$ \\
\hline
$M_{\rho'(1450)}$ [GeV/$c^2$]      & $1.35$ (fixed) & --- & --- \\
$\Gamma_{\rho'(1450)}$ [GeV]       & $0.32$ (fixed) & --- & --- \\
$A^R_{\rho'(1450)}$                & $0.0043 \pm 0.0024$ & --- & --- \\
$A^I_{\rho'(1450)}$                & $0.0213 \pm 0.0038$ & --- & --- \\
\hline
$M_{f_2}$ [GeV/$c^2$]              & --- & $1.2754$ (fixed) & $1.2754$ (fixed) \\
$\Gamma_{f_2}$ [GeV]               & --- & $0.1858$ (fixed) & $0.1858$ (fixed) \\
$A^R_{f_2}$                        & --- & $0.0386 \pm 0.0130$ & $0.0000 \pm 0.0000$ \\
$A^I_{f_2}$                        & --- & $0.0000 \pm 0.0000$ & $0.0386 \pm 0.0130$ \\
\hline
$B^R_{\gamma A}$                   & $-0.3693 \pm 0.0056$ & $-0.3633 \pm 0.0050$ & $-0.3633 \pm 0.0050$ \\
\hline
$C^R_{\gamma\gamma}$               & --- & $-0.0516 \pm 0.0199$ & $0$ (fixed) \\
$C^I_{\gamma\gamma}$               & --- & $0$ (fixed) & $-0.0516 \pm 0.0199$ \\
\hline
Reduced $\chi^2$                   & $286.0/157 = 1.82$ & $288.6/157 = 1.84$ & $288.6/157 = 1.84$ \\
\hline\hline
\end{tabular}
\caption{\textcolor{black}{Extracted fit parameters for the spin-1 (Model~A)
and spin-2 (Model~B) descriptions of the ALICE dipion spectrum. For the
spin-2 model, results are listed for two fixed choices of the overall phase
of the two-photon sector relative to the photonuclear amplitudes, $\phi=0$
and $\phi=\pi/2$.}}
\label{tab:fit_results}
\end{threeparttable}
\end{table}

\section{Fit parameters and phases}
\label{sec:merged-fits}
\added{Table~\ref{tab:fit_results} summarizes the extracted parameters for
the spin-1 (Model~A) and spin-2 (Model~B) fits shown in Fig.~1 of the Letter.
Because interference between $C$-odd (photonuclear) and $C$-even (two-photon)
amplitudes vanishes in the azimuthally integrated invariant-mass spectrum
under pion-exchange-symmetric selections, the mass fit constrains only the
relative phase between the $f_2(1270)$ resonance and the $\gamma\gamma$
continuum, leaving their common phase $\phi$ relative to the photonuclear
sector unconstrained. Allowing the real and imaginary parts of the two-photon
amplitudes to vary independently therefore introduces large parameter
degeneracies. Accordingly, we fix this common phase to $\phi=0$ or
$\phi=\pi/2$ in the spin-2 fits; both choices yield identical invariant-mass
distributions and $\chi^2/\mathrm{ndf}$, while producing the distinct $A_1$
and $A_3$ predictions shown in Fig.~2 of the Letter.}
\color{black}

\section{Azimuthal harmonics}
\label{sec:merged-angular}
\added{For outgoing pion four-momenta $p_1=p_{\pi^+}$ and $p_2=p_{\pi^-}$,
we define the dipion kinematic variables as}
\begin{equation}
 Q^2=(p_1+p_2)^2,\qquad
 q_\perp=p_{1\perp}+p_{2\perp},\qquad
 P_\perp=\frac{p_{1\perp}-p_{2\perp}}{2},\qquad
 \Delta\phi=\phi_{P_\perp}-\phi_{q_\perp}.
 \label{eq:focused-momenta}
\end{equation}
\added{Here hats denote transverse unit vectors, $\tilde b_\perp$ is the
nucleus--nucleus impact parameter, and primes label momenta in the conjugate
amplitude. Adopting the photon-polarization approximation of Eq.~(1) in the
Letter, the coherent $\rho^0(770)$ photoproduction cross section takes the
schematic form}~\cite{xing20}:
\begin{equation}\label{eq:old-rho-schematic}
    \begin{split}
    \frac{d \sigma_{\rho \rightarrow \pi \pi}} {d^3 p_1 d^3 p_{2}d^2  \tilde b_{\perp} }&\sim\frac{1}{2 (2\pi)^7} 
    \frac{P_\perp^2}{(M_{\pi\pi}^2 - M_\rho^2)^2 + M_\rho^2 \Gamma_\rho^2} 
     f_{\rho \pi \pi}^2 \int d^2 \Delta_\perp \, d^2 k_\perp \, d^2 k'_\perp \,
    (\hat P_\perp \cdot \hat k_\perp)(\hat P_\perp \cdot \hat k'_\perp),
    \end{split}
\end{equation}
\added{For clarity, nuclear form factors and impact-parameter phase factors
are suppressed in Eqs.~\eqref{eq:old-rho-schematic} and
\eqref{eq:old-continuum-schematic}; the complete $\rho^0$ amplitude and
resonance-interference expressions are derived in Sec.~\ref{sec:focused-tensor},
where $k_\perp$ and $k'_\perp$ denote the photon transverse momenta in the
amplitude and its complex conjugate, respectively.}

\begingroup
\added{The factor $(\hat P_\perp\cdot\hat k_\perp)(\hat P_\perp\cdot\hat k'_\perp)$
generates a second-harmonic modulation in $\Delta\phi$. Cosine moments
isolate terms that are even under reflection across the pair transverse
momentum $q_\perp$. Provided the phase-space integration and kinematic
selections remain invariant under simultaneous reflection of the pion
momenta, internal transverse momenta, and impact-parameter orientation,
the nuclear factors become independent of the pion azimuth after integration.
Projecting the photon unit vectors onto directions parallel and perpendicular
to $\hat q_\perp$ yields}
\begin{equation}
    \label{pdotk}
    \begin{split}
        (\hat P_\perp\cdot\hat k_\perp)(\hat P_\perp\cdot\hat k'_\perp)
        ={}&\cos^2(\Delta\phi)
        (\hat q_\perp\cdot\hat k_\perp)(\hat q_\perp\cdot\hat k'_\perp)+\sin^2(\Delta\phi)\big[
        \hat k_\perp\cdot\hat k'_\perp
        -(\hat q_\perp\cdot\hat k_\perp)(\hat q_\perp\cdot\hat k'_\perp)
        \big].
    \end{split}
\end{equation}
\added{which, upon applying double-angle identities, reduces to}
\par\endgroup
\begin{equation}
    \label{2phidotprod}
    \begin{split}
        (\hat P_\perp& \cdot \hat k_\perp)(\hat P_\perp \cdot \hat k'_\perp) =\frac12\{\hat k_\perp \cdot \hat k_\perp' + \cos(2\Delta\phi)[2(\hat q_\perp \cdot \hat k_\perp)(\hat q_\perp \cdot \hat k'_\perp ) - \hat k_\perp \cdot \hat k_\perp']\},
    \end{split}
\end{equation}
\added{Similarly, the interference cross section between coherent $\rho^0$
photoproduction and the nonresonant $\gamma\gamma\to\pi^+\pi^-$ continuum
can be written schematically as}~\cite{hagiwara21}
\begin{equation}\label{eq:old-continuum-schematic}
    \begin{split}
        \frac{d \sigma_I} {d^3 p_1 d^3 p_{2}d^2 \tilde b_{\perp} }&\sim \frac{2 M_\rho \Gamma_\rho |P_\perp| f_{\rho\pi\pi}}{(Q^2 - M_\rho^2)^2 + M_\rho^2 \Gamma_\rho^2} \int d^2 \Delta_\perp\, d^2 k_\perp\, d^2 k'_\perp\ \delta^{(2)}(k_\perp + \Delta_\perp - q_\perp)\\&
\times\left[
  \hat{k}_\perp \cdot \hat{\Delta}_\perp
  - \frac{2 P_\perp^2}{P_\perp^2 + m_\pi^2}
    (\hat{k}_\perp \cdot \hat{P}_\perp)
    (\hat{\Delta}_\perp \cdot \hat{P}_\perp)
\right]
(\hat{P}_\perp \cdot \hat{k}'_\perp).
    \end{split}
\end{equation}
\added{Its azimuthal dependence is governed by two scalar-product structures:}
\begin{equation}
    \begin{split}
        &(\hat{k}_\perp \cdot \hat{\Delta}_\perp)(\hat P_\perp \cdot \hat k_\perp'), \\
        &(\hat{k}_\perp \cdot \hat P_\perp)(\hat{\Delta}_\perp \cdot \hat P_\perp)(\hat P_\perp \cdot \hat k_\perp').
    \end{split}
\end{equation}
\added{Projecting onto the reflection-even sector, the linear term yields
a pure first harmonic,}
\begin{equation}
    \label{1phidotprod}
    (\hat{k}_\perp \cdot \hat{\Delta}_\perp) (\hat q_\perp\cdot \hat k_\perp')\cos(\Delta\phi).
\end{equation}
\begingroup
\added{The reflection-even part of the cubic product expands as}
\begin{equation}
    \begin{split}
        &\cos^3(\Delta\phi)
        (\hat q_\perp\cdot\hat k_\perp)
        (\hat q_\perp\cdot\hat k'_\perp)
        (\hat q_\perp\cdot\hat\Delta_\perp)\\
        &+\cos(\Delta\phi)\sin^2(\Delta\phi)\Big[
        (\hat k_\perp\cdot\hat k'_\perp)
        (\hat q_\perp\cdot\hat\Delta_\perp)
        +(\hat q_\perp\cdot\hat k_\perp)
        (\hat k'_\perp\cdot\hat\Delta_\perp)\\
        &\qquad+(\hat q_\perp\cdot\hat k'_\perp)
        (\hat k_\perp\cdot\hat\Delta_\perp)
        -3(\hat q_\perp\cdot\hat k_\perp)
        (\hat q_\perp\cdot\hat k'_\perp)
        (\hat q_\perp\cdot\hat\Delta_\perp)\Big].
    \end{split}
\end{equation}
\added{Applying triple-angle identities separates this contribution into
first and third harmonics:}
\begin{equation}
    \label{3phidotprod}
    \begin{split}
        &(\hat k_\perp\cdot\hat P_\perp)
        (\hat\Delta_\perp\cdot\hat P_\perp)
        (\hat P_\perp\cdot\hat k'_\perp)\\
        &\quad=\frac{\cos(\Delta\phi)}{4}\Big[
        (\hat k_\perp\cdot\hat k'_\perp)
        (\hat q_\perp\cdot\hat\Delta_\perp)
        +(\hat q_\perp\cdot\hat k_\perp)
        (\hat k'_\perp\cdot\hat\Delta_\perp)+(\hat q_\perp\cdot\hat k'_\perp)
        (\hat k_\perp\cdot\hat\Delta_\perp)\Big]\\
        &\qquad+\frac{\cos(3\Delta\phi)}{4}\Big[
        4(\hat q_\perp\cdot\hat k_\perp)
        (\hat q_\perp\cdot\hat k'_\perp)
        (\hat q_\perp\cdot\hat\Delta_\perp)
        -(\hat k_\perp\cdot\hat k'_\perp)
        (\hat q_\perp\cdot\hat\Delta_\perp)\\
        &\qquad\quad-(\hat q_\perp\cdot\hat k_\perp)
        (\hat k'_\perp\cdot\hat\Delta_\perp)
        -(\hat q_\perp\cdot\hat k'_\perp)
        (\hat k_\perp\cdot\hat\Delta_\perp)\Big].
    \end{split}
\end{equation}
\par\endgroup
\added{Equations~\eqref{1phidotprod} and \eqref{3phidotprod} contain exclusively
odd harmonics, which integrate to zero over $\Delta\phi$ for
pion-exchange-symmetric selections. Their observable strengths in $A_1$ and
$A_3$ are determined by the underlying production amplitudes and nuclear
convolution integrals.}

\section{\texorpdfstring{$\rho^0$--$f_2$}{rho0--f2} helicity amplitudes and interference}
\label{sec:focused-tensor}
\added{To go beyond the factorized spin-channel approximation used in the
Letter, we compute the $\rho^0$--$f_2(1270)$ interference directly from the
underlying production amplitudes, including both tensor helicity states,
the full photon transverse-momentum dependence, and the interference
between the two nucleus-emitter configurations.}

\paragraph{Production amplitudes.}
\added{At fixed impact parameter $\tilde b_\perp$, let $\mathcal F(x,k_\perp)$
and $\mathcal A_{\rm co}(x,\Delta_\perp)$ denote the quasireal photon emission
amplitude and the coherent photonuclear scattering amplitude, respectively.
In the small-$q_\perp$ limit, the longitudinal momentum fractions carried
by the colliding fields for pion rapidities $y_{1,2}$ are}
\begin{equation}
 x_1=\sqrt{\frac{P_\perp^2+m_\pi^2}{s_{NN}}}
       (e^{y_1}+e^{y_2}),\qquad
 x_2=\sqrt{\frac{P_\perp^2+m_\pi^2}{s_{NN}}}
       (e^{-y_1}+e^{-y_2}).
 \label{eq:focused-fractions}
\end{equation}
\added{Because the quasireal photon is linearly polarized along
$\hat k_\perp$, coherently summing the two emitter--target amplitudes gives}
\begin{align}
 \mathcal T_\rho(\tilde b_\perp)
 ={}&\int\dd^2k_\perp\,\dd^2\Delta_\perp\,
 \delta^{(2)}(q_\perp-k_\perp-\Delta_\perp)
 \frac{\ii f_{\rho\pi\pi}P_\perp}{D_\rho(Q)}
 (\hat P_\perp\cdot\hat k_\perp)
 \nonumber\\
 &\times\Big[
 e^{-\ii\tilde b_\perp\cdot k_\perp}
 \mathcal F(x_1,k_\perp)\mathcal A_{\rm co}(x_2,\Delta_\perp)
 +e^{-\ii\tilde b_\perp\cdot\Delta_\perp}
 \mathcal F(x_2,k_\perp)\mathcal A_{\rm co}(x_1,\Delta_\perp)
 \Big],
 \label{eq:focused-rho}\\
 D_\rho(Q)={}&Q^2-m_\rho^2+\ii m_\rho\Gamma_\rho.
 \label{eq:focused-rho-denominator}
\end{align}
\added{Here $\Delta_\perp$ is the transverse momentum transfer from the
target nucleus and $f_{\rho\pi\pi}$ is the effective
$\rho^0\to\pi^+\pi^-$ coupling. The corresponding differential cross section is}
\begin{equation}
 \frac{\dd\sigma_{\rho\to\pi\pi}}
 {\dd^2p_{1\perp}\dd^2p_{2\perp}\dd y_1\dd y_2
  \dd^2\tilde b_\perp}
 =\frac{|\mathcal T_\rho(\tilde b_\perp)|^2}{2(2\pi)^7}.
 \label{eq:focused-rho-rate}
\end{equation}

\added{In two-photon fusion $\gamma\gamma\to f_2(1270)\to\pi^+\pi^-$, equal
photon helicities populate the $\lambda=0$ state, whereas opposite photon
helicities populate the $|\lambda|=2$ states. The corresponding helicity
amplitudes are}
\begin{align}
 \mathcal M_{++}^{f_2}=\mathcal M_{--}^{f_2}
 ={}&-\frac{a_{f_2\gamma\gamma}g_{f_2\pi\pi}}
 {M_0D_{f_2}(Q)}
 F_{f_2\pi\pi}(Q^2)F_{f_2\gamma\gamma}(Q^2)
 \frac{Q^4(Q^2-4m_\pi^2)}{12}(3\cos^2\theta-1),
 \label{eq:focused-helicity-zero}\\
 \mathcal M_{+-}^{f_2}=\mathcal M_{-+}^{f_2}
 ={}&-\frac{b_{f_2\gamma\gamma}g_{f_2\pi\pi}}
 {M_0D_{f_2}(Q)}
 F_{f_2\pi\pi}(Q^2)F_{f_2\gamma\gamma}(Q^2)
 \frac{Q^2(Q^2-4m_\pi^2)}{4}\sin^2\theta,
 \label{eq:focused-helicity-two}\\
 D_{f_2}(Q)={}&Q^2-m_{f_2}^2+\ii m_{f_2}\Gamma_{f_2},\qquad
 \sin^2\theta=\frac{4P_\perp^2}{Q^2-4m_\pi^2}.
 \label{eq:focused-tensor-denominator}
\end{align}

\added{The couplings, reference scale $M_0$, and off-shell form factors are
specified in Sec.~\ref{sec:focused-example}. Contracting with the linear
polarization vectors $\hat k_{1\perp}$ and $\hat k_{2\perp}$ yields}
\begin{align}
 \mathcal M_{f_2}
 \equiv\hat k_{1\perp}^{\mu}\hat k_{2\perp}^{\nu}
       \mathcal M_{\mu\nu}^{f_2}
 ={}&\left[2(\hat k_{1\perp}\cdot\hat P_\perp)
 (\hat k_{2\perp}\cdot\hat P_\perp)
 -\hat k_{1\perp}\cdot\hat k_{2\perp}\right]
 \mathcal M_{+-}^{f_2}
 \nonumber\\
 &-(\hat k_{1\perp}\cdot\hat k_{2\perp})\mathcal M_{++}^{f_2}.
 \label{eq:focused-polarization}
\end{align}
Equations~\eqref{1phidotprod} and \eqref{3phidotprod} show that,
 helicity $0$ contributes to
the first harmonic, while helicity $2$ contributes to both the
first and third harmonics.

\paragraph{Interference and yield.}
\added{Under pion exchange ($P_\perp\to-P_\perp$,
$\Delta\phi\to\Delta\phi+\pi$), the $C$-odd $\rho^0$ amplitude reverses sign
while the $C$-even two-photon amplitude remains invariant. Consequently, their
interference cancels in the azimuthally integrated yield for
pion-exchange-symmetric cuts but survives in odd azimuthal harmonics---a
property that holds equally for resonant and nonresonant $C$-even two-photon
amplitudes.}

\added{For a generic $C$-even resonance $R\in\{f_0,f_2\}$, the
$\rho^0$--$R$ interference cross section is}
\begin{align}
 &\frac{\dd\sigma_{\rm int}^{(R)}}
 {\dd^2p_{1\perp}\dd^2p_{2\perp}\dd y_1\dd y_2
  \dd^2\tilde b_\perp}
 =-\frac{1}{Q^2(2\pi)^5}
 \frac{P_\perp f_{\rho\pi\pi}}{\sqrt{4\pi}}
 \mathcal K_{\rho R}(Q)
 \int\dd^2k_\perp\,\dd^2\Delta_\perp\,\dd^2k_\perp'
 \nonumber\\
 &\quad\times\delta^{(2)}(q_\perp-k_\perp-\Delta_\perp)
 (\hat P_\perp\cdot\hat k_\perp')
 \Big\{-(\hat k_\perp\cdot\hat\Delta_\perp)
 D_R(Q)\mathcal M_{++}^{R}
 \nonumber\\
 &\qquad\quad+
 [2(\hat k_\perp\cdot\hat P_\perp)
    (\hat\Delta_\perp\cdot\hat P_\perp)
   -\hat k_\perp\cdot\hat\Delta_\perp]
 D_R(Q)\mathcal M_{+-}^{R}\Big\}
 \nonumber\\
 &\quad\times\Big\{
 e^{\ii\tilde b_\perp\cdot(k_\perp'-k_\perp)}
 \mathcal F(x_1,k_\perp)\mathcal F(x_2,\Delta_\perp)
 \mathcal F^*(x_1,k_\perp')
 \mathcal A_{\rm co}^*(x_2,\Delta_\perp')
 \nonumber\\
 &\qquad\quad+
 e^{\ii\tilde b_\perp\cdot(\Delta_\perp'-k_\perp)}
 \mathcal F(x_1,k_\perp)\mathcal F(x_2,\Delta_\perp)
 \mathcal F^*(x_2,k_\perp')
 \mathcal A_{\rm co}^*(x_1,\Delta_\perp')\Big\}.
 \label{eq:focused-interference}
\end{align}
\added{Here the propagator phase factor is defined as}
\begin{equation}
 \mathcal K_{\rho R}(Q)
 =\Imm\!\left[\frac{1}{D_\rho(Q)D_R^*(Q)}\right]
 =\frac{(Q^2-m_\rho^2)\Imm D_R(Q)
 -m_\rho\Gamma_\rho\Ree D_R(Q)}
 {|D_\rho(Q)|^2|D_R(Q)|^2},\qquad R=f_0,f_2.
 \label{eq:focused-phase}
\end{equation}

\added{The scalar $f_0(980)$ amplitudes are detailed in
Sec.~\ref{sec:focused-scalar}.}

\added{Setting $k_{2\perp}=q_\perp-k_{1\perp}$ and
$k'_{2\perp}=q_\perp-k'_{1\perp}$, the pure two-photon $f_2(1270)$
cross section reads}
\begin{align}
 &\frac{\dd\sigma_{f_2}}
 {\dd^2p_{1\perp}\dd^2p_{2\perp}\dd y_1\dd y_2
  \dd^2\tilde b_\perp}
 =\frac{1}{16\pi^2Q^4}
 \int\dd^2k_{1\perp}\,\dd^2k_{2\perp}
 \frac{\dd^2k_{1\perp}'}{(2\pi)^2}
 \delta^{(2)}(q_\perp-k_{1\perp}-k_{2\perp})
 \nonumber\\
 &\quad\times e^{\ii(k_{1\perp}-k_{1\perp}')\cdot\tilde b_\perp}
 \Big\{
 (\hat k_{1\perp}\cdot\hat k_{2\perp})
 (\hat k'_{1\perp}\cdot\hat k'_{2\perp})
 |\mathcal M_{++}^{f_2}|^2
 \nonumber\\
 &\qquad+\Big[2(\hat k_{1\perp}\cdot\hat P_\perp)
 (\hat k_{2\perp}\cdot\hat P_\perp)
 -\hat k_{1\perp}\cdot\hat k_{2\perp}\Big]
 \Big[2(\hat k'_{1\perp}\cdot\hat P_\perp)
 (\hat k'_{2\perp}\cdot\hat P_\perp)
 -\hat k'_{1\perp}\cdot\hat k'_{2\perp}\Big]
 |\mathcal M_{+-}^{f_2}|^2
 \nonumber\\
 &\qquad-\Big[2(\hat k_{1\perp}\cdot\hat P_\perp)
 (\hat k_{2\perp}\cdot\hat P_\perp)
 -\hat k_{1\perp}\cdot\hat k_{2\perp}\Big]
 (\hat k'_{1\perp}\cdot\hat k'_{2\perp})
   \mathcal M_{++}^{f_2}\mathcal M_{+-}^{f_2*}
 \nonumber\\
 &\qquad-(\hat k_{1\perp}\cdot\hat k_{2\perp})
 \Big[2(\hat k'_{1\perp}\cdot\hat P_\perp)
 (\hat k'_{2\perp}\cdot\hat P_\perp)
 -\hat k'_{1\perp}\cdot\hat k'_{2\perp}\Big]
   \mathcal M_{+-}^{f_2}\mathcal M_{++}^{f_2*}\Big\}
 \nonumber\\
 &\quad\times
 \mathcal F(x_1,k_{1\perp})\mathcal F^*(x_1,k_{1\perp}')
 \mathcal F(x_2,k_{2\perp})\mathcal F^*(x_2,k_{2\perp}').
 \label{eq:focused-tensor-rate}
\end{align}
\added{For the azimuthally symmetric fiducial selections adopted in
Sec.~\ref{sec:focused-example}, the cross terms between the helicity-$0$ and
$|\lambda|=2$ amplitudes in Eq.~\eqref{eq:focused-tensor-rate} vanish upon
integrating over the pion azimuth.}

\section{Numerical \texorpdfstring{$\rho^0$--$f_2$}{rho0--f2} results}
\label{sec:focused-example}

\added{We now evaluate the $\rho^0+f_2(1270)$ model numerically for
neutron-inclusive Pb+Pb UPCs at LHC energies using the formalism of
Sec.~\ref{sec:focused-tensor}.}

\subsection{Pb+Pb inputs}
\label{sec:pbpb-setup}
\added{Our numerical calculations consider Pb+Pb collisions ($A=208$,
$Z=82$) at $\sqrt{s_{NN}}=5.02~\mathrm{TeV}$ within the fiducial window
$1.0<Q<1.3~\mathrm{GeV}$, $|Y_{\pi\pi}|<0.8$, and $|\eta_{1,2}|<0.9$.
Events are selected inclusively in forward-neutron emission, and all
predictions are evaluated at the particle level without detector effects.}

\begingroup\color{black}
The photon amplitude and nuclear charge form factor~{\color{black}\cite{Shao:2022stc}} are
\begin{align}
 \mathcal F(x,k_\perp)
 &=\frac{Z\sqrt{\alpha_{\rm em}}}{\pi}
 \frac{|k_\perp|}{k_\perp^2+x^2m_p^2}
 F_A(k_\perp^2+x^2m_p^2),
 \label{eq:focused-photon}\\
 F_A(q^2)&=\frac{3}{q^3R_c^3}
 [\sin(qR_c)-qR_c\cos(qR_c)]\frac{1}{1+a^2q^2},
 \qquad R_c=1.1A^{1/3}~\mathrm{fm},\quad a=0.7~\mathrm{fm}.
 \label{eq:focused-charge}
\end{align}
\added{Here $q$ in $F_A(q^2)$ denotes the three-momentum transfer magnitude
(distinct from the pair transverse momentum $q_\perp$). For the
$\rho^0(770)$, we adopt $m_\rho=0.775~\mathrm{GeV}$,
$\Gamma_\rho=0.147~\mathrm{GeV}$, and $f_{\rho\pi\pi}=11.87$. The
$f_2(1270)$ resonance parameters and couplings are fixed to the values of
Ref.}~\cite{Bolz:2014mya}:
\begin{align}
 m_{f_2}&=1.275~\mathrm{GeV},\qquad
 \Gamma_{f_2}=0.185~\mathrm{GeV},\qquad M_0=1~\mathrm{GeV},
 \label{eq:focused-tensor-mass}\\
 g_{f_2\pi\pi}&=9.26,\qquad
 a_{f_2\gamma\gamma}=\frac{e^2}{4\pi}(1.45~\mathrm{GeV}^{-3}),\qquad
 b_{f_2\gamma\gamma}=\frac{e^2}{4\pi}(2.49~\mathrm{GeV}^{-1}).
 \label{eq:focused-couplings}
\end{align}
\added{Off-shell effects are incorporated via the exponential form factors}
\begin{equation}
 F_{f_2\pi\pi}(Q^2)=F_{f_2\gamma\gamma}(Q^2)
 =\exp\!\left[-\frac{Q^2-m_{f_2}^2}{\Lambda_{f_2}^2}\right],
 \qquad \Lambda_{f_2}=1.0,\ 1.5~\mathrm{GeV}.
 \label{eq:focused-form-factors}
\end{equation}

\added{The two cutoffs $\Lambda_{f_2}=1.0$ and $1.5~\mathrm{GeV}$ serve to
quantify the theoretical sensitivity to the off-shell prescription. For
neutron-inclusive UPCs, the impact-parameter magnitude $|\tilde b_\perp|$
is integrated from $2R$ to $\infty$.}
\par\endgroup

\subsection{Predictions: phase, helicity, and form factors}
\label{sec:pbpb-sensitivity}
\begin{figure}[!t]

 \centering
 \includegraphics[width=\textwidth]{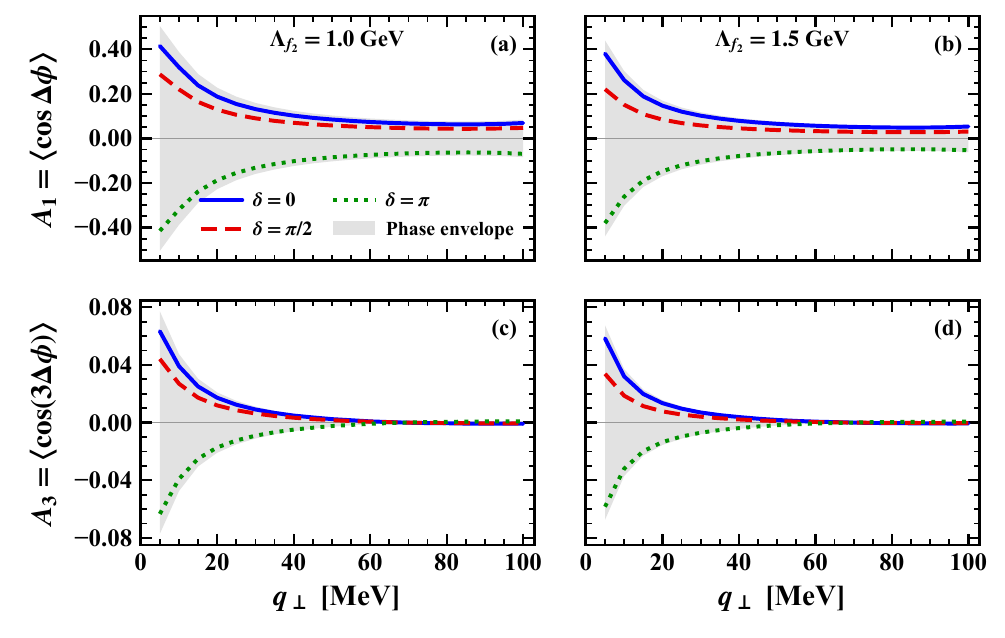}
 \caption{\textcolor{black}{Predicted $A_1$ and $A_3$ in the
 $\rho^0+f_2(1270)$ model for neutron-inclusive Pb+Pb UPCs at
 $\sqrt{s_{NN}}=5.02~\mathrm{TeV}$, including both tensor helicity states.
 Kinematic cuts are $1.0<Q<1.3~\mathrm{GeV}$, $|Y_{\pi\pi}|<0.8$, and
 $|\eta_{1,2}|<0.9$ in the leading small-$q_\perp$ approximation. Left and
 right panels correspond to $\Lambda_{f_2}=1.0$ and $1.5~\mathrm{GeV}$,
 respectively. Curves display $\delta=0,\pi/2,\pi$, and the shaded bands
 indicate the full envelope spanned by varying $\delta\in[0,2\pi)$ at
 each $q_\perp$.}}
 \label{fig:pbpb-phase}
\end{figure}

\added{To assess the sensitivity of $A_1$ and $A_3$ to the relative
production phase between the $\rho^0$ and $f_2(1270)$ channels, we multiply
both tensor helicity amplitudes by $e^{\ii\delta}$, with $\delta=0$
corresponding to the baseline phase convention of
Eq.~\eqref{eq:focused-phase}. Because pion-exchange symmetry eliminates
the interference term from the azimuthally integrated cross section, the
denominator of $A_n$ is independent of $\delta$, and the odd harmonics vary
harmonically with $\delta$:}
\begin{equation}
 A_n(q_\perp;\delta)
 =A_n(q_\perp;0)\cos\delta
 +A_n(q_\perp;\pi/2)\sin\delta,\qquad n=1,3.
 \label{eq:pbpb-phase-scan}
\end{equation}
\added{Figure~\ref{fig:pbpb-phase} displays the resulting asymmetries for
$\delta=0,\pi/2,\pi$ together with the phase envelope
$\pm\sqrt{A_n(q_\perp;0)^2+A_n(q_\perp;\pi/2)^2}$. Varying $\delta$
substantially modifies both the magnitude and sign of $A_1$ and $A_3$
across $q_\perp$.}

\begin{figure}[!t]

 \centering
 \includegraphics[width=\textwidth]{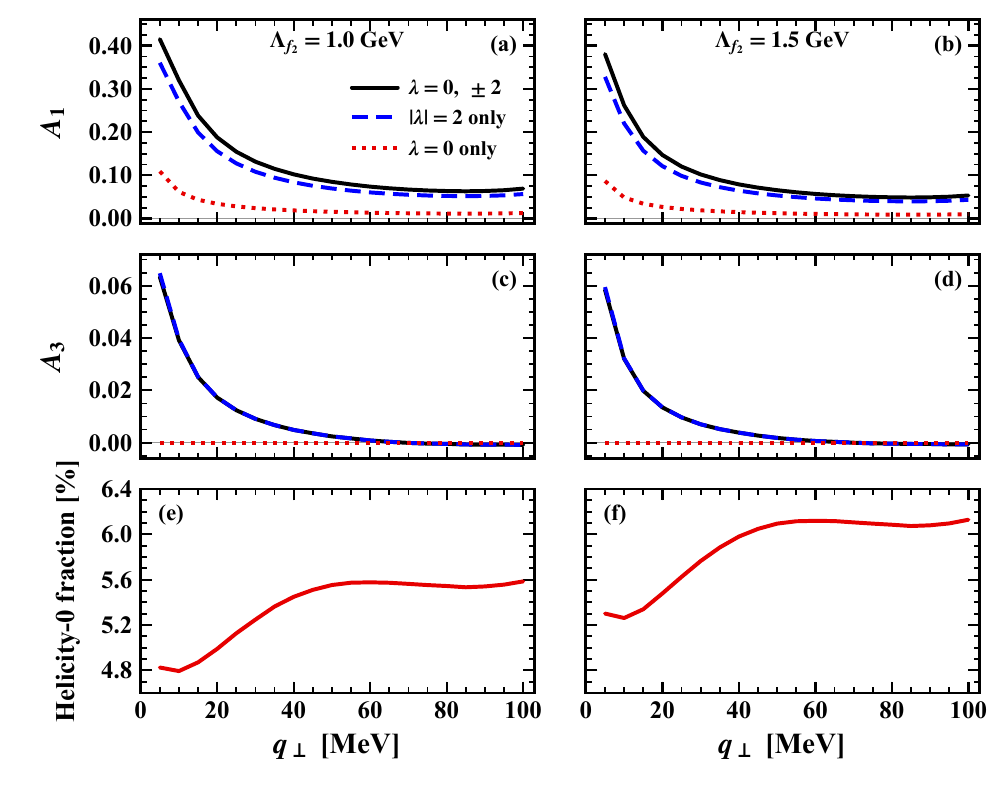}
 \caption{\textcolor{black}{Tensor helicity decomposition of $A_1$ and $A_3$
 at $\delta=0$ for the Pb+Pb kinematic selection of Fig.~\ref{fig:pbpb-phase}.
 Top and middle rows compare the full calculation against the individual
 $|\lambda|=2$ and $\lambda=0$ contributions; the bottom row shows the
 fractional contribution of helicity~$0$ to the pure $f_2(1270)$ yield.}}
 \label{fig:pbpb-helicity}
\end{figure}

\added{Figure~\ref{fig:pbpb-helicity} isolates the tensor helicity
contributions at $\delta=0$ by comparing the full calculation with the
$|\lambda|=2$-only and $\lambda=0$-only cases (applying the same helicity
selection consistently to both the interference numerator and the pure
$f_2$ denominator). Although the helicity-$0$ state accounts for only
$4.8\%$--$6.2\%$ of the pure $f_2(1270)$ cross section across the displayed
$q_\perp$ range, its interference with the $\rho^0$ amplitude is
disproportionately significant: omitting helicity~$0$ reduces $A_1$ at
$q_\perp=5~\mathrm{MeV}$ by $13.1\%$ for $\Lambda_{f_2}=1.0~\mathrm{GeV}$
and by $13.9\%$ for $\Lambda_{f_2}=1.5~\mathrm{GeV}$.}

\begin{figure}[!ht]

 \centering
 \includegraphics[width=0.60\textwidth]{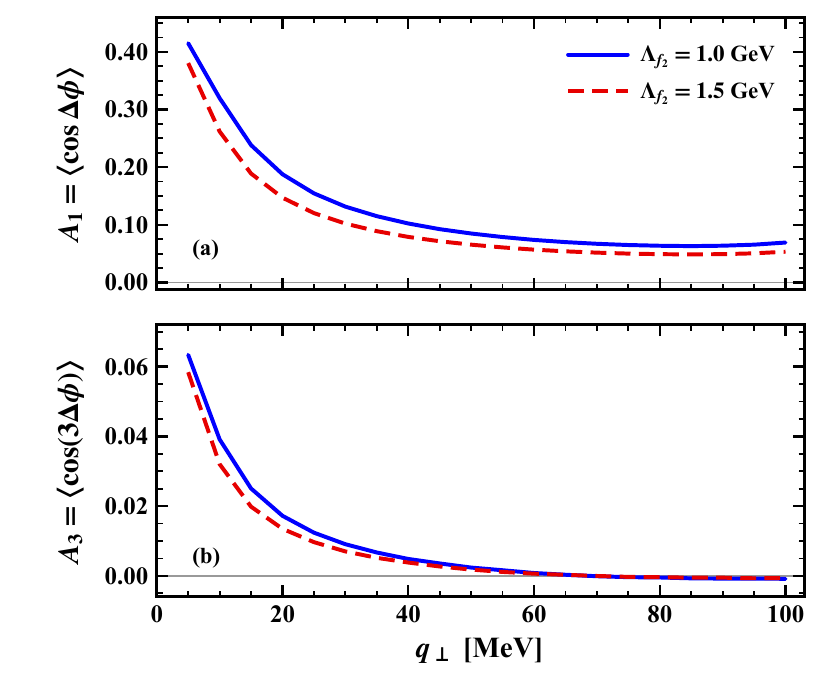}
 \caption{\textcolor{black}{Predicted $A_1$ (top) and $A_3$ (bottom) from
 $\rho^0$--$f_2(1270)$ interference in neutron-inclusive Pb+Pb UPCs at
 $\sqrt{s_{NN}}=5.02~\mathrm{TeV}$ for $\delta=0$, including both tensor
 helicities. Cuts: $1.0<Q<1.3~\mathrm{GeV}$, $|Y_{\pi\pi}|<0.8$, and
 $|\eta_{1,2}|<0.9$. Solid blue and dashed red curves correspond to
 $\Lambda_{f_2}=1.0$ and $1.5~\mathrm{GeV}$, respectively.}}
 \label{fig:focused-qt}
\end{figure}

\added{Figure~\ref{fig:focused-qt} directly compares the predicted
$q_\perp$ dependence of $A_1$ and $A_3$ for $\Lambda_{f_2}=1.0$ and
$1.5~\mathrm{GeV}$ at the reference phase $\delta=0$, using the Pb+Pb setup
of Sec.~\ref{sec:pbpb-setup} with both tensor helicities included.}

\section{Probing the $f_0(980)$ parameters through $A_1$}
\label{sec:focused-scalar}

\added{Interference between coherent $\rho^0$ photoproduction and
$\gamma\gamma\to f_0(980)\to\pi^+\pi^-$ offers a complementary probe of
the scalar resonance through the first harmonic $A_1$. A spin-$0$ state
couples exclusively to equal-helicity photons, giving}
\begin{equation}
 \mathcal M_{++}^{f_0}=\mathcal M_{--}^{f_0}
 =\frac{g_{f_0\gamma\gamma}g_{f_0\pi\pi}}{D_{f_0}(Q)},\qquad
 \mathcal M_{+-}^{f_0}=\mathcal M_{-+}^{f_0}=0.
 \label{eq:focused-scalar-helicity}
\end{equation}
\added{These amplitudes contribute solely to $A_1$ (and not to $A_3$).
To capture the coupled-channel dynamics near the $K\bar K$ threshold, we
adopt the Flatt\'e parametrization:}
\begin{align}
 D_{f_0}(Q)&=Q^2-m_{f_0}^2
 +\ii m_{f_0}[\Gamma_\pi(Q)+\Gamma_K(Q)],
 \label{eq:focused-scalar-denominator}\\
 \Gamma_\pi(Q)&=g_1\sqrt{\frac{Q^2}{4}-m_\pi^2},\qquad
 \Gamma_K(Q)=\frac{g_2}{2}\left[
 \sqrt{\frac{Q^2}{4}-m_{K^+}^2}
 +\sqrt{\frac{Q^2}{4}-m_{K^0}^2}\right],
 \label{eq:focused-scalar-widths}\\
 g_{f_0\pi\pi}&=m_{f_0}\sqrt{\frac{16\pi g_1}{3}}.
 \label{eq:focused-scalar-coupling}
\end{align}
\added{Here $g_1$ and $g_2$ are dimensionless coupling constants, and below
each kaon threshold the square root is analytically continued via
$\sqrt{Q^2/4-m_K^2}\to\ii\sqrt{m_K^2-Q^2/4}$.}
Through $\mathcal K_{\rho f_0}$ in
Eq.~\eqref{eq:focused-interference}, $A_1$ depends on the
$f_0(980)$ line shape and its interference with $\rho^0$. Thus, measurements of $A_1$ could provide information on
parameters associated with the $f_0(980)$ contribution,
including its decay widths and its production phase
relative to $\rho^0$.
\section{Nonresonant two-photon continuum}
\label{sec:nonresonant-continuum}
\begingroup
\added{For comparison with the resonant $\rho^0$--$f_2(1270)$ signal,
Fig.~\ref{fig:rho-born-continuum-final} shows the odd azimuthal asymmetries
generated by interference between coherent $\rho^0$ photoproduction and
the nonresonant Born $\gamma\gamma\to\pi^+\pi^-$ continuum, evaluated within
the framework of Ref.}~\cite{hagiwara21} \added{using the identical Pb+Pb
inputs and fiducial selection as in Sec.~\ref{sec:pbpb-setup}.}

\begin{figure}[!ht]
 \centering
 \includegraphics[width=0.60\textwidth]
 {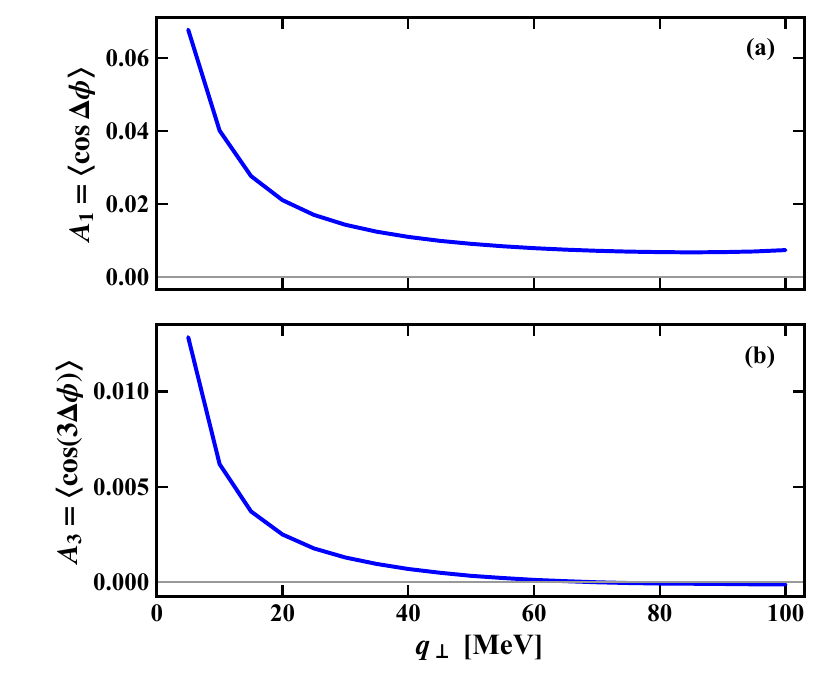}
 \caption{\textcolor{black}{Predicted $A_1=\langle\cos\Delta\phi\rangle$
 (top) and $A_3=\langle\cos(3\Delta\phi)\rangle$ (bottom) arising from
 interference between coherent $\rho^0$ photoproduction and the nonresonant
 Born $\gamma\gamma\to\pi^+\pi^-$ continuum in neutron-inclusive Pb+Pb UPCs
 at $\sqrt{s_{NN}}=5.02~\mathrm{TeV}$, for $1.0<Q<1.3~\mathrm{GeV}$,
 $|Y_{\pi\pi}|<0.8$, and $|\eta_{1,2}|<0.9$.}}
 \label{fig:rho-born-continuum-final}
\end{figure}
\par\endgroup

\color{black}
\bibliographystyle{unsrt}
\bibliography{references}

\makeatletter
\if@filesw\immediate\write\@auxout{\gdef\string\@extra@binfo{}}\fi
\makeatother
\end{document}